\documentclass[conference]{IEEEtran}
\IEEEoverridecommandlockouts
\usepackage{cite}
\usepackage{amsmath,amssymb,amsfonts}
\usepackage{algorithmic}
\usepackage{graphicx}
\usepackage{textcomp}
\usepackage{xcolor}
\usepackage{color, colortbl}
\usepackage{multirow}
\usepackage{flushend}
\usepackage{hyperref}

\newcommand{\MyPara}[1]{\vspace{.2em}\noindent\textit{\textbf{#1}}\hspace{.3em}}
\newcommand{\MyBox}[1]{\vspace{3mm}\noindent\framebox[\columnwidth][c]{\parbox[b]{0.95\columnwidth}{ #1 }}\vspace{3mm}}

\def\BibTeX{{\rm B\kern-.05em{\sc i\kern-.025em b}\kern-.08em
    T\kern-.1667em\lower.7ex\hbox{E}\kern-.125emX}}
\begin{document}

\title{Software Engineering Research Community Viewpoints on Rapid Reviews}



\author{\IEEEauthorblockN{\textsuperscript{1}Bruno Cartaxo, \textsuperscript{2}Gustavo Pinto, \textsuperscript{3}Baldoino Fonseca, \textsuperscript{3}M\'{a}rcio Ribeiro, \\\textsuperscript{3}Pedro Pinheiro,\textsuperscript{4}Sergio Soares, \textsuperscript{6}Maria Teresa Baldassarre}
\IEEEauthorblockA{
\textsuperscript{1} Federal Institute of Pernambuco (IFPE) - Paulista, Pernambuco, Brazil\\
\textsuperscript{2} Federal University of Par\'{a} (UFPA) - Bel\'{e}m, Par\'{a}, Brazil\\
\textsuperscript{3} Federal University of Alagoas (UFAL) - Macei\'{o}, Alagoas, Brazil\\
\textsuperscript{4} Federal University of Pernambuco (UFPE) - Recife, Pernambuco, Brazil\\
\textsuperscript{6} University of Bari (UNIBA) - Bari, Italy\\
email@brunocartaxo.com, gpinto@upfa.br, marcio@ic.ufal.br, baldoino@ic.ufal.br,\\ pmop@ic.ufal.br, mariateresa.baldassarre@uniba.it, scbs@cin.ufpe.br
}
 }

\maketitle

\begin{abstract}
\noindent 
Background: One of the most important current challenges of Software Engineering (SE) research is to provide relevant evidence to practice. In health related fields, Rapid Reviews (RRs) have shown to be an effective method to achieve that goal. However, little is known about how the SE research community perceives the potential applicability of RRs.\\
Aims: The goal of this study is to understand the SE research community viewpoints towards the use of RRs as a means to provide evidence to practitioners.\\
Method:  To understand their viewpoints, we invited 37 researchers to analyze 50 opinion statements about RRs, and rate them according to what extent they agree with each statement. Q-Methodology was employed to identify the most salient viewpoints, represented by the so called \emph{factors}.\\
Results: Four factors were identified: Factor A groups undecided researchers that need more evidence before using RRs; Researchers grouped in Factor B are generally positive about RRs, but highlight the need to define minimum standards; Factor C researchers are more skeptical and reinforce the importance of high quality evidence; Researchers aligned to Factor D have a pragmatic point of view, considering RRs can be applied based on the context and constraints faced by practitioners.\\
Conclusions: In conclusion, although there are opposing viewpoints, there are also some common grounds. For example, all viewpoints agree that both RRs and Systematic Reviews can be poorly or well conducted.
\end{abstract}

\begin{IEEEkeywords}
Rapid Reviews, Systematic Reviews, Q-Methodology, Evidence Based Software Engineering
\end{IEEEkeywords}

\section{Introduction}

Software Engineering (SE) research community has long recognized the role of empirical methods. Part of these advances follow from the sequence of studies conducted by Barbara Kitchenham and colleagues, highly influenced by the medical field, who advocated in favor of the Evidence Based Software Engineering (EBSE) paradigm~\cite{Kitchenham2004,Kitchenham06,BudgenCTBKL06,KitchenhamBBL05,DybaKJ05}. In their seminal work~\cite{Kitchenham2004}, when authors coined the EBSE term, they stated the goal of EBSE is to integrate the best evidence from research with practical experience~\cite{Kitchenham2004}. 

Since then, Systematic Reviews (SRs) became one of the most well-known products of EBSE. The thorough procedure that one conducting a SR should follow made the outcomes more fair, rigorous, and auditable. Due to these benefits, there is no need to say that SE community eagerly embraced EBSE in general, and SRs, in particular~\cite{Borges2015,Borges2014,daSilva2011,Zhou2015}. 
There is a gotcha, however. A SR does not end when all existing evidence is aggregated; a SR is also intended to support practitioners to use the curated knowledge to deal with their SE problems. Therefore, guidelines and implications to practitioners are a crucial part of a SR. However, after these 10+ years from the first introduction of SRs, investigations have shown SRs still lack connection with SE practice. To illustrate, in a survey with 44 authors of 120 SRs, only six of them affirmed their work had direct impact on industrial practice~\cite{Santos2013}.

Another critical issue is related to time and effort needed to conduct SRs, taking from months to one year (sometimes even more) as well as demanding a joint effort of many researchers. This long collaborative effort, although invaluable for curating knowledge, is hardly acceptable for those that require empirical answers in a timely fashion, which is often the case of the SE practitioners~\cite{Cartaxo2018ease}.

This concern pertains not only to the SE community, since the medicine field has also faced a similar problem in its early days~\cite{Best1997,tricco2017rapid,Tricco2015}. One of the most successful mitigation initiatives is what has been called as \textit{Rapid Reviews} (RRs)~\cite{Tricco2015}. They lightweight secondary studies intended to provide evidence to support informed decision making. RRs, are conducted in time frames more likely to meet practitioners' demands (in weeks or in a month, instead of months or a year) and demand the involvement of few, sometimes only one researcher, reducing the costs to deliver evidence.

Although RRs are a rising research method in the medical domain, they are barely known by the SE community. To the best of our knowledge, our recent work on RRs \cite{Cartaxo2018ease} is the only one reporting experiences of applying RRs in SE. In that study we observed that practitioners were very supportive towards the use of RRs. The importance of exploring the perceptions of practitioners, as we have done recently \cite{Cartaxo2018ease}, is easy to understand since practitioners are the target audience of RRs. But the perceptions of researchers should certainly not be neglected since according to Rogers, the perceptions of all individuals involved in an initiative is one of the main predictors of its adoption~\cite{rogers2003diffusion}.

Unfortunately, the perceptions of researchers, who usually are in charge of conducting the RRs, is still unknown. The RRs methodological concessions provoked controversies in the medical field. Considering the recent movements in SE to increase SRs quality~\cite{Zhou2015,niazi2015systematic,dyba2007applying,Hassler2014,Cruzes2011}, we believe RRs could provoke similar controversy in SE too. However, according to Rogers, the perceptions of all individuals involved in an initiative is one of the main predictors of its adoption~\cite{rogers2003diffusion}. Thus, the importance of exploring the perceptions of practitioners, as we have done \cite{Cartaxo2018ease}, is easy to understand since practitioners are the target audience of RRs. But the perceptions of researchers should certainly not be neglected.

\MyBox{\textbf{Our problem}: RRs can speed up the knowledge transfer process to practice (and initial results suggest that practitioners appreciate it), but researchers (who are usually in charge of conducting RRs) may be skeptical to adopt RRs due to methodological concessions.}

In this paper, we investigate the viewpoints of SE researchers on the potential use of RRs to support informed decision making in SE practice. We invited a group of SE researchers; asked them to analyze a predefined set of statements regarding RRs, and rank them according to what extent they agree with each statement. To conduct this analysis, we relied on a Q-Methodology strategy. Different than a survey, Q is a mix of qualitative and quantitative methods to understand the subjectivity underlying the viewpoints of a group of people towards a topic.

The main outcome of this paper is a typology constituted by a set of viewpoints (or factors) that reveals the existent perceptions software engineering researchers regarding the use of RRs. The viewpoints we have encountered in this study are:

\begin{itemize}
    \item[\textbf{FA:}] \textbf{Unconvinced:} The SE researcher needs more evidence about Rapid Reviews;
    \item[\textbf{FB:}] \textbf{Enthusiastic:} The SE researcher is willing to conduct RRs, but only if minimum standards are meet;
    \item[\textbf{FC:}] \textbf{Picky:} The SE researcher does not believe that RRs can deliver high quality evidence;
    \item[\textbf{FD:}] \textbf{Pragmatic:} The SE researcher might conduct some RRs in specific scenarios where they are appropriate.
\end{itemize}

Ultimately, with these viewpoints uncovered, the research community can better address the concerns of researchers according to their viewpoints, in order to make RRs more appealing, not only to practitioners, but also to the research community.  Finally, along with this study, we also implemented a tool that one interested in performing a Q-Methodology study could instantiate and use. More details about the tool, as well as its source code and manual in Section~\ref{subec_qsorting_process}.

\section{Background}
\label{sec_background}

In this section we provide additional information regarding RRs (Section~\ref{subsec_background_rapidreviews}), as well as details about the method we applied in this study, the Q-Methodology (Section~\ref{subsec_background_qmethodology}). 

\subsection{Rapid Reviews}
\label{subsec_background_rapidreviews}

RRs are secondary studies that employ strategies to accelerate the traditional SRs process. The goal is to substantially improve the amount of time to gather, analyze, interpret, review, and publish evidence that could benefit practitioners. To achieve this goal, RRs deliberately omit or simplify steps of traditional SRs. For instance, limiting literature search, using just one person to screen studies, skipping formal synthesis, among others~\cite{Cartaxo2018ease,Tricco2015}.

Although RRs present some method-wise variations, they usually share at least these core characteristics~\cite{Cartaxo2018ease,Cartaxo2016}: (1) they are performed in close collaboration with practitioners and bounded to a practical problem; (2) they reduce costs and time of heavyweight methods; (3) they report results through mediums more appealing to practitioners.

RRs may have limitations due to methodological flexibilization. However, in spite of their limitations, the interest in RRs is growing in medicine field \cite{Tricco2015}. For example, the prestigious Systematic Reviews journal recognized RRs as one of the methods composing Evidence-Based Practice~\cite{Moher2015}. That interest may be explained by some benefits of RRs already reported in the medicine scientific literature. For instance, a study observed RRs saved around \$ 3 millions when applied in a hospital \cite{Mcgregor2005}. A survey exploring the use of 15 RRs has shown that, the knowledge producted with 53\% of them were directly applied in practice \cite{hailey2009}. Other studies also have shown the positive impact of RRs in medicine \cite{hailey2000,batten2012,zechmeister2012,Tricco2015,Taylor-Phillips2018,lawani2017five}.

Furthermore, one should not confuse RRs with informal or ad-hoc literature reviews \cite{niazi2015systematic}. RRs follow systematic protocols, although some methodological concessions are made aiming to deliver evidence to practitioners in a more timely manner.

Another important point is that, by any means, RRs were not conceived to replace SRs. Instead, both should coexist. SRs are important to provide rigorous in-depth insights, while RRs focus on easily and quickly transfer established scientific knowledge to practice \cite{Cartaxo2018ist}. 

\subsection{Q-Methodology}
\label{subsec_background_qmethodology}

Q is a methodology belonging to the psychometric spectrum that employs both qualitative and quantitative methods to shed light on complex issues in which human subjectivity is involved \cite{watts2005doing,brown1993primer,brown1980political,good2010introduction}. The methodology focuses on discovering the existent viewpoints about a particular topic. It is typically conducted throughout the following steps~\cite{zabala2016bootstrapping}:

\MyPara{Defining the Concourse:} The concourse is a set of possible opinion statements referring to a particular topic. ``Opinion'' is the key word here, since the statements do not need to be based on facts, but reflect people perceptions. For instance, the statement ``Rapid reviews are quick and dirty systematic reviews'' may sound strong, but if there are researchers thinking this way, it is a valid statement for this study. The concourse has statements representing all different points of view, pertaining to multiple discourses~\cite{zabala2016bootstrapping}. 

\MyPara{Defining the Q-SET:} The Q-SET is the refined concourse (e.g., removing concourse's duplicated statements). It represents the opinion field for the topic under investigation~\cite{kelly2016expediting}. The Q-SET has typically between 40 and 80 opinion statements about the topic under investigation~\cite{paige2016q}.

\MyPara{Defining the P-SET:} The P-SET represents people that participated in the study. It is obtained by strategic sampling, instead of, for instance, random sampling. This is due to the fact that Q-Methodology aims to ensure comprehensiveness and diversity, rather than representativeness or quantity (as occurs with surveys) \cite{armatas2017understanding}. 
The participants (P-SET) rank the statements (Q-SET) according to how they agree with them.

\MyPara{Defining Q-SORT structure:} The Q-SORT is the distribution of statements ranked by a given participant. It represents the participant's point of view about the topic of interest. As we shall see in Figure~\ref{fig_qsort_structure}, the Q-SORT is usually structured as a quasi-normal distribution~\cite{zabala2016bootstrapping}. Q-SORT structure enables participants to rank few statements in the tails --- the ones they agree or disagree the most --- and more statements in the middle of the distribution.

\MyPara{Conducting Q-Sorting process:} With the Q-SORT structure defined, the participants from the P-SET are invited to rank Q-SET's statements according to how they agree or disagree with them following the Q-SORT structure.

\MyPara{Conducting Factor Analysis:} The literature about Q-Methodology usually refers to its quantitative statistical part as factor analysis \cite{watts2005doing}. However, it is indeed an adaptation and extension of it.  While traditional factor analysis is applied to a population of people, with traits as variables, the Q-Methodology applies to a population of traits (i.e. the statements), with people as variables (i.e. the P-SET)~\cite{good2010introduction}. Factor analysis is usually applied to examine the underlying patterns among a large number of variables and, specially, to summarize the information in smaller set of factors or components~\cite{hair2014multivariate}. A factor represents one of the viewpoints about the topic under investigation. More strictly, a factor represents the Q-SORT of a hypothetical participant who would be the best-representative of the participants with similar points of view~\cite{zabala2016bootstrapping}.

\MyPara{Conducting Factor Interpretation:} Afterwards, one has to interpret the Q-SORTs representing each factor in order to provide a rationale that explains the viewpoints associated with them. This is a writing activity where each factor interpretation includes the statements and their scores in that factor; quotes from participants' comments aligned with that factor; as well as demographic information about them. Comparing factors also reveal interesting information about consensuses and dissensions among viewpoints~\cite{armatas2017understanding}.

\section{Method}\label{sec_method}

In this section we discuss how we instantiated the Q-Methodology in this study but before we discuss why we have chosen Q-Methodology in the first place.

\subsection{Why Q-Methodology?}

We are aware that this methodology might be unfamiliar to the reader, raising valid questions such as why Q-Methodology? We took this decision because we aim at discovering what are the diverse viewpoints about RRs in SE (achievable with Q-Methodology), in opposition to having a sample allegedly representing the proportion of those viewpoints in the SE community population (achievable with survey methods). \textbf{Q-Methodology is not a Survey.} Surveys can lead to biases in the responses, e.g., some participants might be more positive than others. Since the Q-SORT structure follows a quasi-normal distribution, the participants are then required to prioritize their perceptions (instead of being mostly positive/neutral/negative). Moreover, Q-Methodology demands a small number of participants while still preserves the statistical significance.

\subsection{Defining the Q-SET}\label{subsec_defining_pset}

As discussed in Section~\ref{subsec_background_qmethodology}, Q-SET is the refined concourse. However, in this study we neither had to define a concourse nor apply any sampling technique to define the Q-SET. Instead, we replicated the Q-SET defined by Kelly et al. \cite{kelly2016expediting}, which explored the perceptions of researchers and practitioners on applying RRs in medical settings~\cite{kelly2016expediting}. After a careful analysis on their Q-SET, we could see it is fully applicable to the SE context, in particular, after small adjustments on statements mentioning medical related terms. The Q-SET of this study is composed by 50 statements listed in Table~\ref{tab_factor_scores}.

\subsection{Defining the P-SET}

Since the goal of this research is to explore the perceptions of SE researchers on potential applicability of RRs to support decision making in practice, the P-SET of this study is composed by SE researchers only (i.e., we do not cover SE practitioners as Kelly's et al study \cite{kelly2016expediting} do). To maximize the diversity of perspectives, we invited researchers who published at least one paper in the 2015--2017 main tracks of ICSE, (the SE flagship conference) and ESEM and EASE (the leading Empirical Software Engineering conferences). 

We started with 1,293 researchers who published research papers in ICSE, ESEM, or EASE in the 2015, 2016, and 2017 editions. These are the researchers we could extract e-mails to establish contact. From that initial amount, 146 e-mails were bounced, and 34 had an out of office automatic reply. A total of 1,113 e-mails were successfully sent. The P-SET of this study comprises 37 participants, a response rate of 3.3\%. This low percentage was expected since the invitation email announces the questionnaire would take around 45 minutes to be completed and demands reading/understanding the basic concepts of RRs. 

Among our participants, 54\% are between 18 and 35 years old (43\% between 36 and 50, and only 3\% are more than 50 years old). Still, 59\% are located in South America, 32\% in Europe, 3\% in North America, 3\% in Asia, and 3\% in Oceania. Regarding their highest degree, 91\% have a PhD, and 9\% have a masters. The majority of them are researchers in an academic position (86\%), 8\% are researchers working in the industry, and 5\% are practitioners. More interesting for this study, 92\% and 70\% have already read and conducted a SR, respectively. Regarding RRs, 46\% of them have heard of RRs before.

\subsection{Defining the Q-SORT Structure}

\begin{figure}[!ht]
\centering
\includegraphics[width=0.3\textwidth]{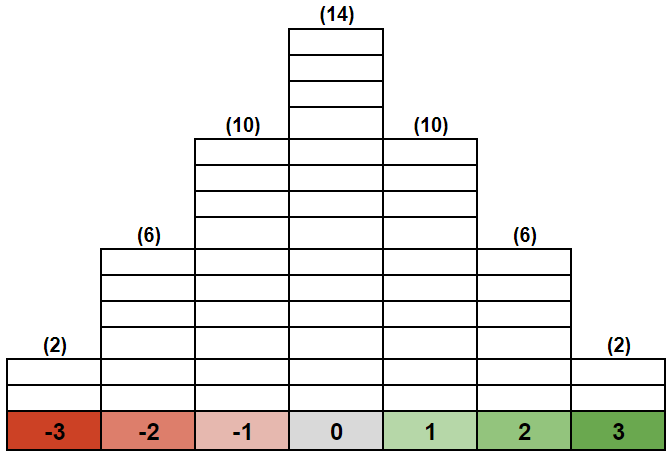}
\caption{Q-SORT structure. The number under parentheses represents the number of statements per column (level of agreement).}
\label{fig_qsort_structure}
\end{figure}

To maintain uniformity and comparability, we followed the same Q-SORT structure as in Kelly's et al. study~\cite{kelly2016expediting}, as depicted in Figure~\ref{fig_qsort_structure}. The Q-SORT structure follows a quasi-normal distribution. It consists of 50 cells (one for each statement) spread across seven columns, representing a rank with seven-point scale from +3 (strongest agreement) to -3 (strongest disagreement). The numbers in parenthesis denote the number of statements to rank per column.

\subsection{Conducting the Q-Sorting process}
\label{subec_qsorting_process}

In the Q-Sorting process, a researcher presents the statements (Q-SET) so participants (P-SET) can rank them according to the predefined Q-SORT structure. This was conducted online in order to reach a wide range of participants. We then implemented a tool to support the Q-Sorting process. This tool is strongly influenced by an existing tool, the HTML-Q (online at: \texttt{\url{https://github.com/aproxima/htmlq}}. After we have faced many bugs in HTML-Q, and have tried to fix them, we ended up concluding it would demand too many structural changes. Thus, we decided to develop another tool from scratch. With the tool fully developed, we conducted a pilot, with four researchers, and got feedback to fix bugs, enhance user experience, as well as to see whether the participants understood the tasks they had to do (e.g., if the RR summary was enough to answer the questions, and how clear the statements are).

\begin{figure}[!ht]
\centering
\includegraphics[width=0.4\textwidth]{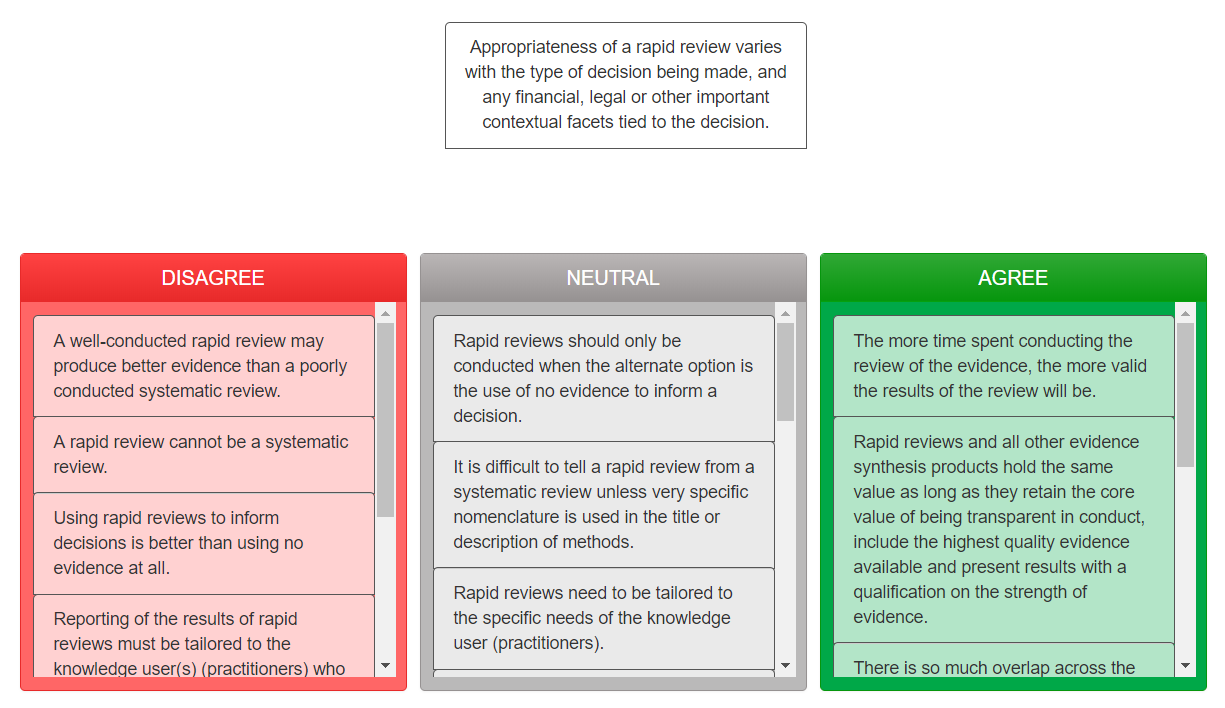}
\caption{3rd step: Classifying statements in three levels of agreement. The white card on top represents the statement the participant should drag and drop to one of the colored boxes. The cards are presented in random order.}
\label{fig_step3}
\end{figure}

The tool enables participants to define their Q-SORTs through six steps. On the first step, participants are introduced to the general information about this study. The second step presents detailed information about RRs, so participants who never heard about it can get acquainted with the general concepts. On the third step, participants are asked to drag and drop 50 cards, each containing one statement of the Q-SET. In this step, participants aimed to classify the cards according to three general levels of agreement --- Agree, Disagree, and Neutral --- as shown in Figure~\ref{fig_step3}. The cards are presented to participants in random order to avoid bias.

\begin{figure}[!ht]
\centering
\includegraphics[width=0.4\textwidth]{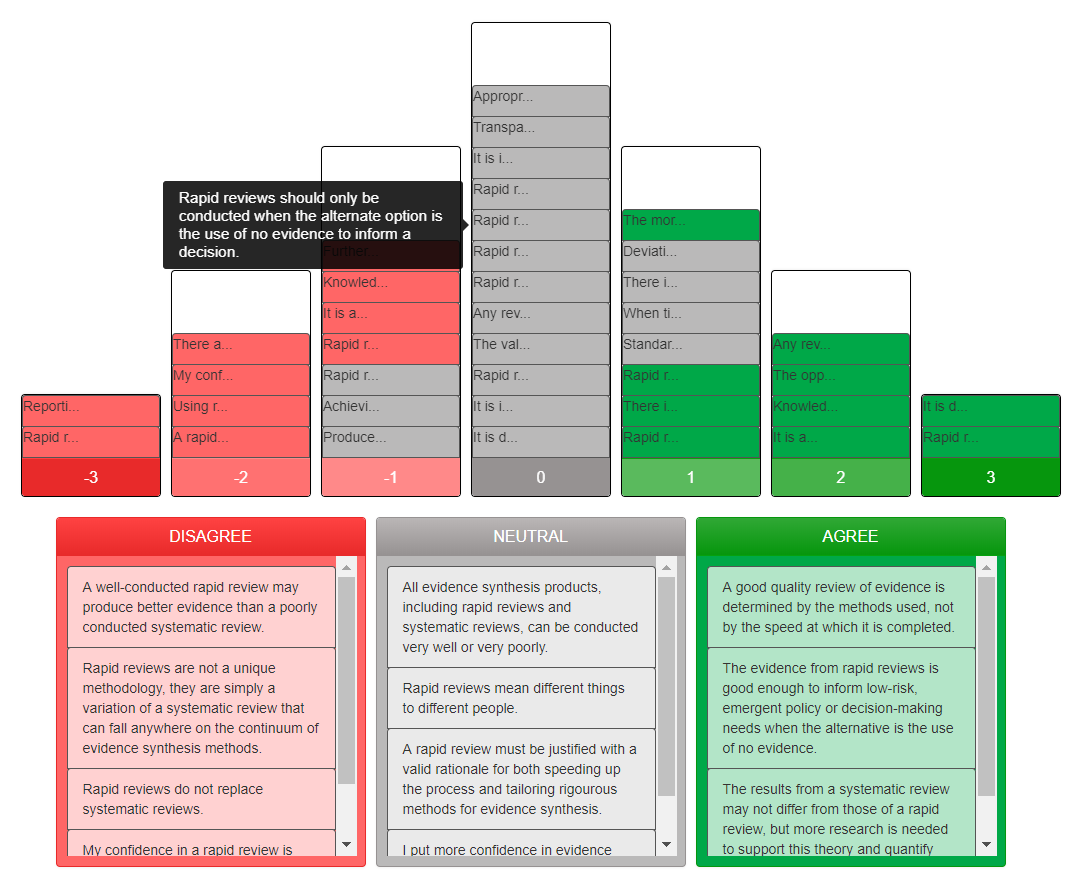}
\caption{4th step: Ranking statements in seven agreement levels according to the Q-SORT structure.}
\label{fig_step4}
\end{figure}

Moving to the fourth step, participants are asked to rate statements in seven levels of agreement in accordance with the Q-SORT structure, as shown in Figure~\ref{fig_step4}.

In the fifth step, participants are asked to explain, why they rated the statements on the extremes (i.e., cards in -3 and +3). We used this qualitative evidence in further analysis during the factor interpretation. On the sixth and last step, participants are asked to answer some demographics questions, and then, finally submit their responses.

For those interested in experimenting our tool, the instance we used in this research is available online at: \texttt{\url{http://qmethod.netlify.com}}. The source code and a manual are also available at \texttt{\url{https://github.com/bfsc/qmethod}}, for anyone interested in instantiating our tool to conduct a different Q-Methodology study.

\subsection{Conducting the Factor Analysis}

\begin{figure}[!ht]
\centering
\includegraphics[width=0.5\textwidth]{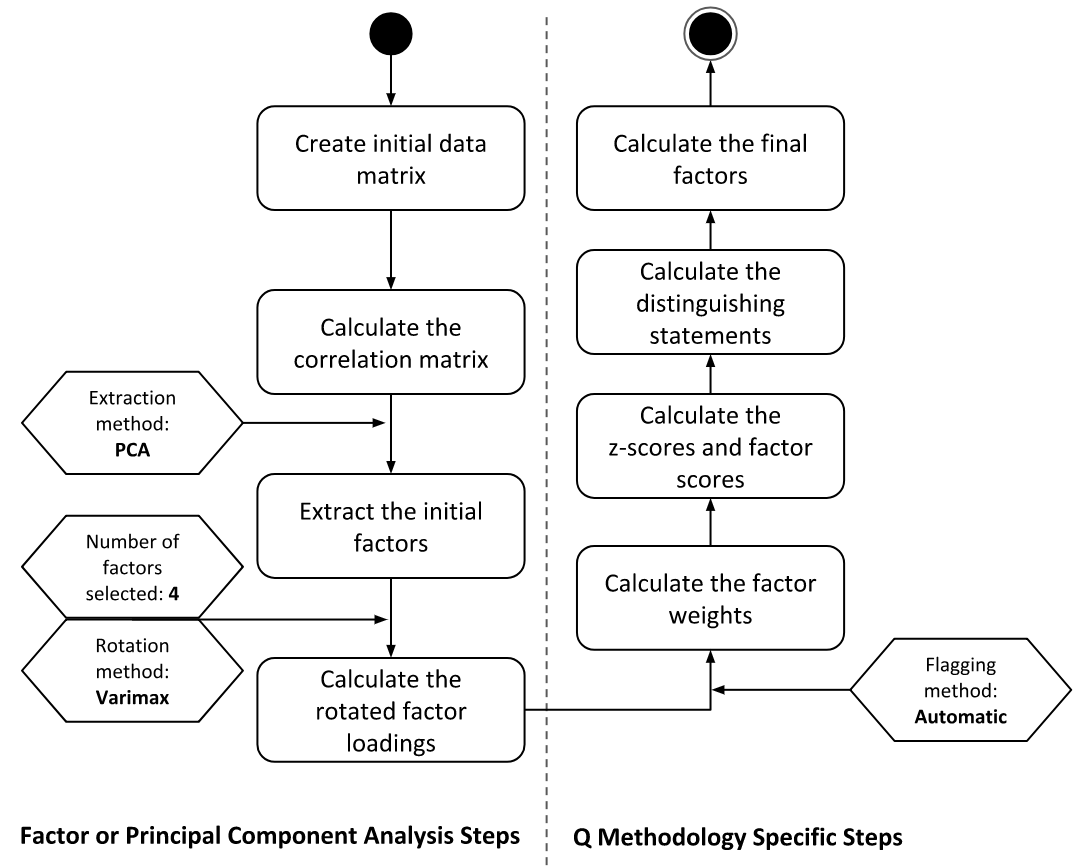}
\caption{Factor analysis steps (rounded rectangles), and decisions made during the analysis (hexagons).}
\label{fig_factor_analysis_steps}
\end{figure}

The Q-SORTs of each of the 37 participants were used as input to the factor analysis. To conduct the analysis we used the \texttt{q-method} R package, which automates part of the data analysis. The text file generated by R as output report for this entire process is available online\footnote{\url{https://bit.ly/2P3JoW5}}. Figure~\ref{fig_factor_analysis_steps} shows the factor analysis steps and the decisions we have made during the process. Following we discuss each step and decisions we made. However, the method is rather complex and explaining all its aspects is out of scope for this paper. One who wants to have better understanding can refer to elsewhere~\cite{watts2005doing,brown1993primer,good2010introduction,zabala2016bootstrapping,paige2016q,hair2014multivariate,abdi2003factor,brown1980political}.

\MyPara{Create initial data matrix:} The Q-SORTs are structured in a two-dimensional matrix (statements $\times$ participants). The value of each cell in this matrix is the value of the column in the Q-SORT (e.g. +3) which the participant placed the statement.

\MyPara{Calculate the correlation matrix:} Next, we correlate the Q-SORTs to create a correlation matrix, which is used as input to extract the initial factors on next step.

\MyPara{Extract the initial factors:} A factor is the weighted average Q-SORT of a group of participants that assigned similar values to the statements, i.e., it represents a hypothetical participant that best represents how those with similar viewpoint would sort the statements~\cite{zabala2016bootstrapping}. We use PCA (Principal Component Analysis) to extract the initial factors. The correlation of each Q-SORTs with each factor is given by the \textit{factor loadings}, which range from −1 to +1. The higher the loading (i.e., the correlation), the more the participant is similar to the factor~\cite{zabala2016bootstrapping}.

\MyPara{Calculate the rotated factor loadings:} Although several factors can be extracted from the Q-SORTs correlation matrix, few factors are capable to explain most of the study's variance. To enhance the explanatory capacity of the factors, a manipulation called \textit{factors rotation} is performed~\cite{abdi2003factor,mulaik2009foundations, brown1993primer,watts2005doing}. Only few factors are selected to be rotated in order to find the optimal arrangement of factor loadings. Quantitative and qualitative criteria are used to define which factors should be selected to be rotated. Some quantitative criteria are extensively used as rule of thumb in literature, such as: (1) when at least two participants load significantly in each factor, which means all factors have at least two participants highly correlated to them; (2) composite reliability greater or equal to 0.8 for each factor; (3) eigenvalues above 1 for each factor, and (4) the percentage of cumulative explained variance among all the selected factors should be between 60\% and 40\%~\cite{kelly2016expediting,zabala2016bootstrapping, watts2005doing, baker2014q, armatas2017understanding}. These criteria indicates the factors strength. Among the initial extracted factors, we selected the first four, which adhered to the quantitative criteria, as shown in Table~\ref{tab_factor_char}. The number of participants (Q-sorts) loaded in the factors are less than the 37 participants. This happens because not all participants could be loaded in a factor (e.g., one participant could have a viewpoint that was not shared among the other participants).

\begin{table}[!ht]
\scriptsize
\begin{center}
\caption{Selected factors characteristics.}
\label{tab_factor_char}

\begin{tabular}{l|c|c|c|c}
\hline
\multirow{2}{*}{\textbf{CHARACTERISTICS}} &  \multicolumn{4}{|c}{\textbf{FACTORS}} \\ \cline{2-5}
& \textbf{A} & \textbf{B} & \textbf{C} & \textbf{D} \\
\hline \hline

Number of loading Q-sorts 			& 12 & 8 & 6 & 4\\ \hline 
Composite reliability 				& 0.98 & 0.97 & 0.96 & 0.94 \\ \hline 
Eigenvalues 						& 6.23 & 5.03 & 3.98 & 3.09 \\ \hline 
Cumulative explained variance 	& 16.84\% & 30.43\% & 41.2\% & 49.55\% \\ \hline 

\end{tabular}
\end{center}
\end{table}

Despite the four factor solution adherence to the above mentioned criteria, we also investigated a range of solutions from three to six factors in detail, all adhering to the quantitative criteria. To choose among the available solution, we applied the qualitative criteria. This leaves researchers free to consider any solution they understand to be the most appropriate and theoretically informative~\cite{zabala2016bootstrapping, watts2005doing}. We excluded the solutions with five and six factors because their extra factors exhibited few or no distinguishing statements. As we shall see in more detail in the next steps, a distinguishing statement is a statement ranked in a factor that significantly differs from its rank in all other factors. On the opposite side, the solution with three factors does not reveal one of the factors presented on the four factor solution, and that factor has high theoretical explanation power. After selecting the number of factors, we performed the rotation applying \textit{Varimax} method \cite{zabala2016bootstrapping}. This rotation resulted in a matrix of \textit{rotated factor loadings} correlating the participant Q-SORTs with the rotated factors.

\MyPara{Calculate the factor weights:} After calculating the rotated factor loadings, it is time to define the final loadings. This is done by \textit{flagging} participants who are the best representatives of each factor. When flagging, one aims at maximizing differences between factors \cite{zabala2016bootstrapping}. We run the automatic flagging, which is based on two criteria: (i) Q-SORTs with factor loading higher than the threshold for p-value $<$ 0.05; (ii) Q-SORTs with square \textit{loading} higher than the sum of square loadings of the same Q-SORT in all other factors \cite{qmethodR}.

\MyPara{Calculate the z-scores and factor scores:} \textit{Z-scores} and \textit{Factor Scores} indicate statements' relative position within each factor. The z-score is a weighted average of the values given by the flagged participants to a statement, and it is continuous. Factor scores are integer values based on z-scores. Z-scores indicate the relationship between statements and factors: how much each factor agrees with a statement. Factor scores are calculated ordering statements by z-score, and matching to the values as established by the Q-SORT structure (i.e. +3 til -3). Factor scores enables us to reconstruct the Q-SORT of a factor, which is crucial to factor interpretation \cite{zabala2016bootstrapping}. Table~\ref{tab_factor_scores} presents the factor scores for each statement of the four factors. It is the output of this step, and crucial to factor interpretation.

\begin{table*}[!ht]
\scriptsize
\begin{center}
\caption{Statements and factor scores. The highlighted cells in \colorbox{green!25}{green} and \colorbox{red!25}{red} indicate the extreme scores of each factor, whereas an asterisk ``*'' is used to denote the distinguishing statements of each factor. Consensus statements are in bold, while dissensions in italic.}
\label{tab_factor_scores}

\begin{tabular}{c|c|c|c|c|c}
\hline
\multirow{2}{*}{\textbf{\#}} &  \multirow{ 2}{*}{\textbf{STATEMENT}} &  \multicolumn{4}{|c}{\textbf{FACTOR SCORES}} \\ \cline{3-6}
&& \textbf{A} & \textbf{B} & \textbf{C} & \textbf{D} \\
\hline \hline

1&\parbox[t]{14cm}{The evidence from rapid reviews is good enough to inform low-risk, emergent policy or decision-making needs when the alternative is the use of no evidence}&0&+1&0&\cellcolor{green!25}+3\\ \hline 
2&\parbox[t]{14cm}{When time allows, a comprehensive systematic review of all available evidence should always be conducted}&-1&-1&+1&\cellcolor{red!25}-3\\ \hline 
3&\parbox[t]{14cm}{Deviating from accepted systematic review methods may introduce bias and impact the validity of the resulting rapid review, which may be an unacceptable risk for some knowledge users (practitioners)}&-1*&0&0&0\\ \hline 
4&\parbox[t]{14cm}{\textit{Further research comparing the methods and results of rapid reviews and systematic reviews is required before I decide how I feel about rapid reviews}}&1&-1&0&-2\\ \hline 
5&\parbox[t]{14cm}{Rapid reviews are too focused in scope and/or context to be generalizable to a variety of knowledge users (practitioners)}&0&-1&-2&0\\ \hline 
6&\parbox[t]{14cm}{\textit{Rapid reviews mean different things to different people}}&0&-1&+1&+2\\ \hline 
7&\parbox[t]{14cm}{Rapid reviews should only precede a more comprehensive and rigorous systematic review}&\cellcolor{red!25}-3&-2&-1*&-2\\ \hline 
8&\parbox[t]{14cm}{The opportunity cost of a comprehensive systematic review is too high and it is more advantageous to conduct rapid reviews when timeliness is a factor}&0&+2&-1&0\\ \hline 
9&\parbox[t]{14cm}{\textit{Rapid reviews do not replace systematic reviews}}&\cellcolor{green!25}+3&+1&+2&0\\ \hline 
10&\parbox[t]{14cm}{\textbf{All evidence synthesis products, including rapid reviews and systematic reviews, can be conducted very well or very poorly}}&\cellcolor{green!25}+3&+2&+2&+2\\ \hline 
11&\parbox[t]{14cm}{Rapid reviews are comparable to systematic reviews except they are done in a more timely fashion}&-2&+1&-1&0\\ \hline 
12&\parbox[t]{14cm}{Rapid reviews are 'quick and dirty' systematic reviews}&-2&\cellcolor{red!25}-3&\cellcolor{red!25}-3&0\\ \hline 
13&\parbox[t]{14cm}{Rapid reviews need to be tailored to the specific needs of the knowledge user (practitioners)}&+1&0&0&+2\\ \hline 
14&\parbox[t]{14cm}{Rapid reviews meet the needs of knowledge users (practitioners)}&+1&+2&0&0\\ \hline 
15&\parbox[t]{14cm}{There are few evidence on rapid reviews, so I cannot support or oppose their use in decision-making}&-1&-2&0&-1\\ \hline 
16&\parbox[t]{14cm}{There is so much overlap across the various evidence synthesis methods that I cannot generalize my opinion to favor one over the other without the context of the decision at hand}&0&0*&+1&+1\\ \hline 
17&\parbox[t]{14cm}{There is a risk involved in tailoring accepted systematic review methods to produce rapid reviews that we do not yet understand}&0&0&0&+1*\\ \hline 
18&\parbox[t]{14cm}{Using rapid reviews to inform decisions is better than using no evidence at all}&+2&+2&-1*&+2\\ \hline 
19&\parbox[t]{14cm}{It is always appropriate to conduct a rapid review}&\cellcolor{red!25}-3&0&-2&-2\\ \hline 
20&\parbox[t]{14cm}{Rapid reviews and all other evidence synthesis products hold the same value as long as they retain the core value of being transparent in conduct, include the highest quality evidence available and present results with a qualification on the strength of evidence}&0&+1&0&-1\\ \hline 
21&\parbox[t]{14cm}{Appropriateness of a rapid review varies with the type of decision being made, and any financial, legal or other important contextual facets tied to the decision}&0&+1&+1&\cellcolor{green!25}+3\\ \hline 
22&\parbox[t]{14cm}{My confidence in a rapid review is impacted by which methods are tailored to speed up the review process}&+1&0&+2&+1\\ \hline 
23&\parbox[t]{14cm}{My confidence in a rapid review is directly tied to results being presented and contextualized by the strength and applicability of the evidence}&0&0&\cellcolor{green!25}+3*&+1\\ \hline 
24&\parbox[t]{14cm}{It is important to have minimum standards for the methodological conduct of rapid reviews}&+2&\cellcolor{green!25}+3&+2&0*\\ \hline 
25&\parbox[t]{14cm}{\textit{It is important to have minimum standards for the reporting of rapid reviews}}&0&\cellcolor{green!25}+3&+2&-1\\ \hline 
26&\parbox[t]{14cm}{Standardization of rapid review methods may conflict with the needs of knowledge users (practitioners)}&-1&-1&0&+1\\ \hline 
27&\parbox[t]{14cm}{The value of rapid reviews in the context of emergent decision-making needs outweighs the disadvantages or risk of bias and potentially 'imperfect' evidence}&-1&+1&-1&+1\\ \hline 
28&\parbox[t]{14cm}{Knowledge users don't always need all of the evidence, they just need the best evidence to support their decision, and what is 'best evidence' is specific to the knowledge user (practitioner)}&+1&-1&-1&-2\\ \hline 
29&\parbox[t]{14cm}{Knowledge users (practitioners) do not fully understand the implications of streamlining evidence synthesis methods to produce a more timely evidence product}&0&-1&+1*&-1\\ \hline 
30&\parbox[t]{14cm}{Reporting of the results of rapid reviews must be tailored to the knowledge user(s) (practitioners) who commissioned the review}&+1&0&+1&+1\\ \hline 
31&\parbox[t]{14cm}{Rapid reviews that omit an assessment of the quality of included studies are useless to knowledge users (practitioners)}&-2&-1&\cellcolor{green!25}+3&0\\ \hline 
32&\parbox[t]{14cm}{Rapid reviews can be timely and valid, even when methodological concessions are made}&+2&+1&-2*&+1\\ \hline 
33&\parbox[t]{14cm}{Transparency of process is more important than the actual methods used to produce rapid reviews, as transparency allows the knowledge user (practitioners) to make their own assessment on validity and appropriateness}&-1&0&0&+2\\ \hline 
34&\parbox[t]{14cm}{It is appropriate to endeavor to define a single, unique methodology for rapid reviews}&0&+1&\cellcolor{red!25}-3&\cellcolor{red!25}-3\\ \hline 
35&\parbox[t]{14cm}{Rapid reviews are not a unique methodology, they are simply a variation of a systematic review that can fall anywhere on the continuum of evidence synthesis methods}&+1&0&0&0\\ \hline 
36&\parbox[t]{14cm}{The results from a systematic review may not differ from those of a rapid review, but more research is needed to support this theory and quantify why results may be the same or different}&+1&+1&+1&-1*\\ \hline 
37&\parbox[t]{14cm}{\textit{I put more confidence in evidence produced in a systematic review than of a rapid review}}&2&-2&0&-1\\ \hline 
38&\parbox[t]{14cm}{\textbf{The more time spent conducting the review of the evidence, the more valid the results of the review will be}}&-2&\cellcolor{red!25}-3&-2&-2\\ \hline 
39&\parbox[t]{14cm}{Achieving a precise estimate of effect (from a systematic review) may not inform the decision-at-hand any better than a general estimate of effect (produced by a rapid review)}&-2&0&-2&0\\ \hline 
40&\parbox[t]{14cm}{Rapid reviews should only be conducted when the alternate option is the use of no evidence to inform a decision}&-2&-2&0*&-2\\ \hline 
41&\parbox[t]{14cm}{A well-conducted rapid review may produce better evidence than a poorly conducted systematic review}&+1&+2&+1&+1\\ \hline 
42&\parbox[t]{14cm}{Any review of evidence that takes longer than one month to produce is not a rapid review}&-1&-2&-1&+2*\\ \hline 
43&\parbox[t]{14cm}{Any review of evidence that takes longer than two weeks to produce is not a rapid review}&-1&0&-2&+1\\ \hline 
44&\parbox[t]{14cm}{A rapid review must be justified with a valid rationale for both speeding up the process and tailoring rigorous methods for evidence synthesis}&+2&+1&+1&-1*\\ \hline 
45&\parbox[t]{14cm}{A good quality review of evidence is determined by the methods used, not by the speed at which it is completed}&0&+2&+2&0\\ \hline 
46&\parbox[t]{14cm}{It is difficult to tell a rapid review from a systematic review unless very specific nomenclature is used in the title or description of methods}&-1&0&-1&-1\\ \hline 
47&\parbox[t]{14cm}{A rapid review cannot be a systematic review}&0&-1&-1&-1\\ \hline 
48&\parbox[t]{14cm}{Rapid review' is too broad a phrase—doing a review in a more timely way can only be relative to how long it takes the same team to produce a systematic literature review}&-1&-1&+1*&-1\\ \hline 
49&\parbox[t]{14cm}{Producers (researchers) are more concerned with the methodology and validity of rapid reviews than knowledge users (practitioners)}&+2&0&-1&0\\ \hline 
50&\parbox[t]{14cm}{It is difficult to judge the validity of a rapid review as the reporting is often truncated and protocols are not published}&+1&-2&0&0\\ \hline 

\end{tabular}
\end{center}
\end{table*}

\MyPara{Calculate the distinguishing statements:} For each factor, we identify their \textit{distinguishing statements}. If a statement ranks in a position that significantly differs from its rank in all other factors, it is distinguishing for that factor. Thus, distinguishing statements are key to interpret factors, as they emphasize factors differences \cite{zabala2016bootstrapping}. The threshold for a difference to be considered significant is calculated based on the Standard Errors for Differences (SED) and the confidence level, which in the case of this study it is 0.05 \cite{zabala2016bootstrapping}. If the difference in z-scores is larger than the threshold, then the statement distinguishes one factor from all others.

\MyPara{Calculate the final factors:} At this step, final factors are achieved. Every statement has its own factor scores, as shown in Table~\ref{tab_factor_scores}. The next step is the factor interpretation.

\section{Results}
\label{sec_results}

To interpret the factors we adhered to the following strategy. First, we analyze the statements on the extremes of the Q-SORT representing each factor (i.e., the ones ranked with -3 or +3). Then, we analyzed the distinguishing statements of each factor. This approach helped us to perceive nuances of each viewpoint that could not be understood just looking to the extremes. During the factor interpretation procedure, two researchers worked together to interpret Factor A. Afterwards, one researcher interpreted Factors B, C, and D, and another researcher revised the process. Finally, we analyzed the participants' comments explaining why they chose to rank the statements they ranked on the extremes (i.e., -3 and +3). Those comments gave additional qualitative evidence that helped us build coherent viewpoints.

Throughout this section, we present the statement scores in two forms. The first when we want to show the statement score for the current factor only. For instance, (S1: +3), which means for the factor under analysis, the statement number 1 is ranked as +3. The second form, we use when we want to highlight the difference between how the factor under analysis scores a statement, compared to the others. For instance, [S4: \textbf{+1*}, -1, 0, -2], which means we are analyzing the Factor A (in bold), and in that factor the statement number 4 is ranked as +1, while in factor B its is ranked as -1, in Factor C as 0, and in Factor D as -2. The asterisk in Factor's A score means the statement number 4 is a distinguishing statement of that factor. 

\subsection{Factors Interpretation}

\MyBox{FACTOR A: \textbf{I'm unconvinced; show me more evidence about Rapid Reviews.}}

Researchers aligned with this viewpoint are undecided; they need more evidence about the benefits/challenges regarding using RRs.

\MyPara{Demographic Characteristics:} Twelve participants are loaded in this factor, which corresponds to 32\% of all participants, and  explains 16.84\% of the total variance of this study. They are 42 years old on average, and ten of them hold a PhD degree. Eleven have conducted a SR themselves, but only one has heard about RRs before.

\MyPara{Viewpoint Attitudes and Perceptions:} ``Further research comparing the methods and results of rapid reviews and systematic literature reviews is required before I decide how I feel about rapid reviews'' [S4: \textbf{+1*}, -1, 0, -2].  As one participant commented, \textit{``We need studies contrasting the precision/recall/efficacy of alternative review methods to help us understand the benefits/problems of using these methods''}. Only on this factor, the participants agreed with that statement, indicating they are undecided concerning the use of RRs. The indecision of this viewpoint towards RRs is even more explicit when we look the contradictory affirmations the participants provided. They think ``A well-conducted rapid review may produce better evidence than a poorly conducted systematic review'' (S42: +1), but on the other hand, they have more confidence in evidence produced with a SR than in evidence produced with a RR (S37: +2). Similarly, the participants strongly disagree that ``It is always appropriate to conduct a rapid review'' (S19: -3), but they agree that ``Rapid reviews can be timely and valid, even when methodological concessions are made.''(S32: +2).  

\MyBox{FACTOR B: \textbf{I'm enthusiastic; Let's conduct rapid reviews, but with minimum standards.}}

Researchers aligned with this viewpoint are generally favorable about RRs, and believe RRs can provide reasonable evidence to practitioners, if minimum standards are established.

\MyPara{Demographic Characteristics:} Eight participants are loaded in this factor, which corresponds to 21\% of all participants, and explains 13.59\% of the total variance of this study. They are 37 years old in average, and six of them hold a PhD degree. The majority has conducted a SR themselves, six out of the eight, and three had heard about RRs before.

\MyPara{Viewpoint Attitudes and Perceptions:} Minimum standards to conduct (S24: +3) and report (S25: +3) RRs are essential. As a consequence, ``Rapid reviews meet the needs of knowledge users (practitioners)'' (S14: +2). 
The general positive perception of this viewpoint can be understood when comparing SRs with RRs. The participants think that ``All evidence synthesis products, including rapid reviews and systematic reviews, can be conducted very well or very poorly'' (S10: +2). Another statement corroborating with this reasoning is: ``A well-conducted rapid review may produce better evidence than a poorly conducted systematic review'' (S41: +2). 
While it was perceived that ``Using rapid reviews to inform decisions is better than using no evidence at all'' (S18: +2), researchers with this viewpoint believe RRs have more to offer, than just when there is no evidence at all. This can be seen when they strongly disagree that ``Rapid reviews should only be conducted when the alternate option is the use of no evidence to inform a decision'' (S40: -2), as well as when they believe RRs have life on its own, and as such, strongly disagree that ``Rapid reviews should only precede a more comprehensive and rigorous systematic review'' (S7: -2).

The comment of a participant explained why s/he is strongly against the idea that ``Rapid reviews are 'quick and dirty' systematic reviews'' (S12: -3) captured the essence of this viewpoint: \textit{``Just because simple reviews can simplify a few steps from systematic reviews does not mean that the results produced have less quality or are wrong. Therefore, I also agree that RRs should have a minimum standardization to avoid such judgments.''}

\MyBox{FACTOR C: \textbf{I'm picky; Rapid reviews might not deliver high quality evidence.}}

Researchers aligned with this viewpoint are very concerned about (1) the quality of primary studies included in RRs and (2) how the primary studies are reported. They also tend to avoid RRs in favor of SRs as much as possible, although they admit they could use RRs in very strict situations.

\MyPara{Demographic Characteristics:} Six participants are loaded in this factor, which corresponds to 16\% of all participants, and explains 10.77\% of the total variance of this study. They are 33 years old on average, and just one do not hold a PhD degree. Only two have conducted a SR themselves, but all have read a SR before. Regarding RRs, two are aware of their existence.

\MyPara{Viewpoint Attitudes and Perceptions:} It was strongly perceived that ``Confidence in a rapid review is directly tied to results being presented and contextualized by the strength and applicability of the evidence'' (S23: +3). This statement is aligned with the perception that ``Rapid reviews that omit an assessment of the quality of included studies are useless to knowledge users (practitioners)'' (S31: +3). Concerns related to evidence quality can be understood if we look to the underlying belief of researchers with this viewpoint. It is considered that ``Knowledge users (practitioners) do not fully understand the implications of streamlining evidence synthesis methods to produce a more timely evidence product'' [S29: 0,-1,\textbf{+1*},-1]. Participants loaded in this factor slightly agree with that statement, but this agreement is particularly interesting when compared to scores of other factors. It points out that participants aligned with the viewpoint of Factor C are the only ones that believe practitioners are not able to understand the implications of potentially low quality evidence.

Researchers with this viewpoint also put little faith in RRs validity. They strongly disregard the possibility that ``Rapid reviews can be timely and valid, even when methodological concessions are made'' [S32: +2, +1, \textbf{-2}, +1], which again significantly distinguishes this viewpoint from the other factors. The strictness of this viewpoint on using RRs is perceived when looking at the factor scores of three distinguishing statements: ``Rapid reviews should only be conducted when the alternate option is the use of no evidence to inform a decision'' [S40: -2, -2, \textbf{0*}, -2], ``Using rapid reviews to inform decisions is better than using no evidence at all'' [S18: +2, +2, \textbf{-1*}, +2]. Despite the strong concern about RRs, a participant loaded on this factor commented that \textit{``Rapid reviews can still be valuable in some contexts''} when asked why s/he ranked the statement ``Rapid reviews are 'quick and dirty' systematic reviews'' (S12) with a -3 score.

\MyBox{FACTOR D: \textbf{I'm pragmatic; let's conduct rapid reviews when they are proper.}} 

Researchers aligned with this viewpoint pragmatically focus on variety of contextual information to decide if RRs are the best fit to support decision-making. They also believe practitioners are able to understand the impacts of flexible research methods adopted by RRs. Still, they believe rigid standards in RRs could reduce their usefulness to practitioners.

\MyPara{Demographic Characteristics:} Four participants are loaded in this factor, which corresponds to 10\% of all participants, and explains 8.35\% of the total variance of this study. They are 36 years old on average, and all of them hold a PhD degree. Two have conducted a SR themselves, but all have read a SR before. Regarding RRs, two are aware of their existence.

\MyPara{Viewpoint Attitudes and Perceptions:} ``Appropriateness of a rapid review varies with the type of decision being made, and any financial, legal or other important contextual facets tied to the decision'' (S21: +3). Thus, for instance, ``The evidence from rapid reviews is good enough to inform low-risk, emergent policy or decision-making needs when the alternative is the use of no evidence'' (S1 : +3). 

Researchers with this viewpoint think RRs should be practice-oriented; they strongly agree that ``Rapid reviews need to be tailored to the specific needs of the knowledge user (practitioners)'' (S13: +2). This practical orientation can also be captured by the following distinguishing statement, ``Any review of evidence that takes longer than one month to produce is not a rapid review'' [S42: -1, -2, -1, \textbf{2*}]. This viewpoint is the only one highlighting the importance of having evidence-based methods in compliance with the well-known time constraints practitioners are usually subject to.

Participants loaded in this factor also believe ``Transparency of process is more important than the actual methods used to produce rapid reviews, as transparency allows the knowledge user (practitioners) to make their own assessment on validity and appropriateness'' (S33: +2). This underlying belief on practitioners ability to do their own assessment about evidence produced with RRs might explain the faith on RRs that researchers with this viewpoint have. However, the faith on RRs does not blurry the researchers loaded in this factor, which corroborates their pragmatic position. This can be seen when they strongly disagree that ``It is always appropriate to conduct a rapid review.'' (S19: -2).

Regarding standardization initiatives, those aligned to this viewpoint strongly disagree that ``It is appropriate to endeavor to define a single, unique methodology for rapid reviews'' (S34: -3). They also are the ones that scored the lowest in the following distinguishing statement ``It is important to have minimum standards for the methodological conduct of rapid reviews.'''[S24: +2, +3, +2, \textbf{0*}]. This pragmatic viewpoint can be better understood with the following comment of a participant about why s/he scored -3 to S34: \textit{``If the purpose of a rapid review is to offer insights to SE practitioners, the methodology will likely depend on the kind of practitioners. I guess the decision makings for safety-critical systems, embedded systems, distributed systems, etc. are different, depending on the kinds of requirements linked to the specific SE sub-field (requiring more or less confidence level, etc.), and one unique methodology is probably not suited for all of them. Plus, SE and research history shows that one unique approach to fit them all has never worked.''}

\subsection{Consensuses and Dissensions Among Factors}

The Consensus Statements reveal the common ground among all the viewpoints~\cite{zabala2016bootstrapping,kelly2016expediting}. 
As depicted in Table~\ref{tab_factor_scores}, participants equally agreed in S10, and equally disagreed in S38.  These statements are indistinguishable across all factors. In particular, all viewpoints believe both RRs and SRs can be conducted very well or very poorly (S10), and time needed to conduct an evidence synthesis study is not related to its quality (S38).

In the opposite direction, participants disagreed when scoring five statements (S37, S25, S9, S6 and S4) making those statements distinguishing across all factors. Controversy among all viewpoints are observable when it is related to topics like: comparing the confidence on evidence produced by SRs and RRs (S37); the importance on defining standards to RRs (S25); if RRs are intended to replace systematic reviews (S9); the meaning of the term RRs (S6); and the need for evidence comparing SRs and RRs (S4). These are referred as Dissension Statements since they reveal controversy among all the viewpoints. 

\section{Discussion}
\label{sec_discussion}

\subsection{Overall Assessment}
Looking at the four factors, we can separate them in two main groups. Factors in the first group (Factors B and D) are more positive or pragmatic towards RRs. On the other hand, factors in the second group (Factors A and C) tend to be undecided or more negative towards RRs. Among many other indicatives, this can be particularly observed looking to one of the controversial statements: ``I put more confidence in evidence produced in a systematic review than of a rapid review''[S37: +2, -2, 0, -1]. It is easy to see that Factors A and C are more confident with SRs, agreeing or being neutral with such statement. However, Factors B and D put more faith in RRs than the other factors since they are on the negative spectrum of agreement regarding that statement.

The proneness to embrace RRs seems to be related to how well researchers know about them. The factors in the first group, which participants are more positive towards RRs (Factors B and D), tend to be more aware of RRs. In factor B, for example, two out of four participants had heard about RRs before. On the opposite side, in the second group, where fewer participants heard about RRs (Factors A and C), the participants tend to be undecided or more negative towards RRs. In Factor A, for example, only one out of 12 participants have heard about RRs before. It might reveal that evaluate the effectiveness of RRs is crucial to convince researchers about RRs applicability. This is particularly supported when we examine one of the controversial statements: ``Further research comparing the methods and results of rapid reviews and systematic reviews is required before I decide how I feel about rapid reviews'' [S4: +1, -1, 0, -2]. While the factors less aware or negative about RRs (A and C) are on the positive or neutral spectrum, demanding more evidence, the factors more aware about RRs (B and D) are on the negative spectrum, disagreeing that more evidence is needed.

\subsection{Researchers' concerns related to Rapid Reviews}

Looking at each factor individually, it is possible to note the main concerns about RRs, although those concerns sometimes are not shared by all viewpoints. Knowing those concerns might be useful to address them, and in turn make RRs more appealing, not only to practitioners, but also to the research community.

\MyPara{More evidence about RRs:} Researchers aligned with Factor A demand more evidence about RRs (S4: +1) but this is not particularly true for the ones aligned with Factor D (S4: -2).

\MyPara{Minimum standards to RRs:} Researchers more aligned with Factor B are more positive about RRs, but strongly believe that minimum standards to conduct (S24: +3) and report (S25: +3) RRs are fundamental. However, again this concern is somewhat controversial, since researchers aligned with Factor D believe that by defining rigid standards to RRs it would make them hard to adapt to the dynamic environment of software practice.

\MyPara{Quality assessment of studies included in RRs:} Researchers aligned to Factor C put little faith on practitioners capacity to assess evidence quality produced with RRs, as well as believe that RRs that skip the quality assessment of primary studies are useless (S31: +3).

\MyPara{Transparency with RRs results:} Researchers aligned with Factor D highlighted the importance of transparency when reporting RRs results so the practitioners can assess the applicability of RRs in their own context. 

\subsection{Viewpoints on Rapid Reviews in Software Engineering compared to Medicine}

By comparing the results of this study with the one conducted in medicine \cite{kelly2016expediting}, it is possible to see some commonalities and distinctions. Both identified a pragmatic viewpoint, as well as a more positive and a more negative ones. In this study however, we have one additional factor, the undecideds. We believe this additional factor emerges because RRs is now well known in medicine, while in SE there is just one study \cite{Cartaxo2018ease} reporting such approach. 

\section{Related Work}
\label{sec_related_work}


Recent studies suggested that the gap between research and practice in SE seems to be more due to the way we conduct research, rather than the topics we investigate~\cite{Lo2015,Carver2016,Begel2014}. To illustrate, Lo and colleagues~\cite{Lo2015} summarized 571 papers from five years of ICSE and ESEC/FSE conferences and asked 512 Microsoft software engineers to rate the studies according to their relevance. 71\% considered the studies as essential or worthwhile. Carver and colleagues conducted an analogous study, but focusing on the ESEM community. Similar results were found: 67\% of all ratings were essential or worthwhile~\cite{Carver2016}. This shows that we, the SE research community, are producing relevant empirical evidence, but we have to find a way to introduce the studies and results in a way that fits practitioners constraints. RRs have shown to be on of the viable solutions \cite{Cartaxo2018ease}. On a different direction, Badampudi and Wohlin have recently proposed a framework to translate knowledge produced through scientific methods but targeting practitioners \cite{Badampudi2019}. Badampudi's work is complementary to RRs since it focuses on translating existent knowledge to practitioners context, while RRs are focused on creating knowledge to be transferred.

Regarding RRs, to the best of our knowledge, there is only the work of Cartaxo et al. \cite{Cartaxo2018ease} approaching the applicability of RRs in SE practice. In this study, a RR was conducted together with practitioners aiming to identify evidence that may support practitioners to mitigate a problem related to low customer collaboration. Practitioners mentioned that by conducting the RR they learned new concepts, improved their comprehension about the problem, and increased their own confidence. Still, they found that RRs offer content that is more reliable than the ones they use to consume to make decisions, and also that the RR process is problem-oriented and applicable to SE practice. 

Finally, only one software engineering study employed Q-Methodology so far~\cite{Winter:2018:ESEM}, which aimed to understand the viewpoints of human values in software engineering. However, besides medicine, other research fields such as agriculture~\cite{davies2007exploring}, ecology~\cite{armatas2017understanding}, etc, are taking advantage of this research method. 

\section{Limitations and Threats to Validity}
\label{sec_limitations}

Probably, the main limitation of this study is that fact that we could not apply a full strategic sampling. This would be the ideal situation because we could deliberatly select some researchers we know have a specific point of view about RRs. We have to resort to a quasi-ramdon sample, otherwise there will be not enough participants. On the other side, some of the most well-known voices of EBSE have participated on this research, which enables us to capture their relevant, and sometimes opposing viewpoints, about RRs. Stil, we cannot strongly support that claim, because to do so, we would need to reveal participants names, hurting their confidentiality. And also due to the presence of participants selected by chance (altough from a set of researchers that published on the most proeminent SE conferences). Thus, we would like to keep highlighting this threat to validity.

Following we list some characteristics of this study that may look threats to validity at first sight for ones not used to Q-Methoodlogy, but indeed are not.

Moreover, in this work we asked the participants to rank their perception regarding RRs. However, 54\% of the participants have never heard about RRs. This may look a threat to validity, but we believe it is not. We provided them a short summary of what is RR and we also could observe that the majority (\texttildelow92\%) have read SRs before. Thus, those participants who are familiar with SRs but have nerver heard about RRs are interesting data points. With them we can capture positive, negative or neutral inclination about RRs regardless participants familiarity with RRs (i.e. prejudices), but safeguarding that they are at least familiar with SRs.

Also, one might consider the number of participants (P-SET) small, in particular, when comparing to survey studies. However, our P-SET is more than three times larger than the one of the medicine study~\cite{kelly2016expediting}. It is also near the range considered as ideal in a Q-Methodology study: between 40 and 60 participants~\cite{watts2005doing}. And as mentioned before, this study aim at discovering what are the diverse viewpoints about RRs in SE (achievable with Q-Methodology), in opposition to having a sample allegedly representing the proportion of those viewpoints in the SE community population (achievable with surveys with large and representative samples).

Yet, one might argue that quasi-normal distribution is not appropriated to translate the participants perceptions regarding the statements asked (e.g., one participant could have a more overall negative perception regarding the use of RRs), therefore, this participant's perception (i.e., the Q-SORT), would more likely follow a negative skewed curve. Although we concur, we believe the mitigation plan here is to have a comprehensive set of statements that cover a wide range of perceptions. Therefore, it could be hardly the case of participants do not agreeing, or agreeing to every statement since many of them are nearly the opposite.

\section{Conclusions}
\label{sec_conclusions}

In this paper we have investigated the SE researchers perceptions and attitudes on RRs. Despite the fact that RRs target practitioners, understanding researchers perceptions is also important since they actively participate on the entire process. Applying a Q-Methodology approach, we revealed four main viewpoints constituting a typology. One is \textbf{unconvinced} and demands more evidence. Another is \textbf{enthusiastic} but believes that establishing minimum standards is important. Other is \textbf{picky} and highlights the importance of quality of evidence. The last one is \textbf{pragmatic} and thinks in terms of applying RRs when the situation is favorable. Despite the differences, we also identified some consensus. For instance, all viewpoints agree that both RRs and traditional SRs can be conducted very well or very poorly. With this typology in mind, one can better understand what are the main concerns of researchers and promote better understanding about RRs. As consequence, we can pave a road better connecting SE research with practice and make SE research community more impacful and relevant.

\vspace{0.2cm}
\emph{Acknowledgments.} We thank the reviewers for their helpful comments and the participants to their willingness to collaborate in this research. 
This work is partially supported by INES (National Institute of Software Engineering), CNPq (National Council for Scientific and Technological Development of the Brazilian government) grants \#465614/2014-0, \#406308/2016-0,  \#304499/2016-1, \#427787/2018-1, and FACEPE (The Foundation for Science and Technology of Pernambuco State) grants APQ-0399-1.03/17 and PRONEX APQ/0388-1.03/14.

\bibliographystyle{abbrv}
\bibliography{references}

\end{document}